\documentclass[usenatbib,twocolumn]{aastex631}

\usepackage{amsmath}	
\usepackage{xcolor}
\usepackage{booktabs}
\usepackage{threeparttable}
\usepackage{comment}
\usepackage{multirow}
\usepackage{algpseudocode}

\makeatletter
\newcounter{algorithm}

\makeatother

\newcommand{\kms}{{\rm km}\,{\rm s}^{-1}}

\newcommand{\msolar}{{\rm M}_{\odot}}

\newcommand{\fgas}{f_{\rm gas}}

\begin{document}

\title{The Deep Learning Halo Definer: A Multimodal Framework for Halo Mass and Gas Fraction Inference on Galaxy Groups and Clusters}

\author[0009-0003-9247-8617]{Caleb Ogle}
\thanks{Email: cbogle@uwm.edu}
\affiliation{Department of Physics and Astronomy, University of Wisconsin-Milwaukee, Milwaukee, WI 53211, USA \\}

\author[0009-0005-7596-4204]{Kalvyn N. Poncelet Adams}
\affiliation{University of California, Los Angeles, Department of Physics and Astronomy, Box 951547, Los Angeles, CA 90095, USA}

\author[0000-0002-3391-2116]{Benjamin D. Oppenheimer}
\affiliation{University of Colorado, Center for Astrophysics and Space Astronomy, 389 UCB, Boulder, CO 80309, USA}

\author[0009-0003-4100-4710]{Naomi Gluck}
\affiliation{Department of Physics, Yale University, New Haven, CT 06520, USA}

\author[0000-0003-3207-8868]{Matthew Ho}
\affiliation{Sorbonne Universit\'{e}, CNRS, UMR 7095, Institut d'Astrophysique de Paris, 98 bis bd Arago, 75014 Paris, France\\}
\affiliation{Department of Astronomy, Columbia University, New York, NY, 10027}

\author[0000-0002-6766-5942]{Daisuke Nagai}
\affiliation{Department of Physics, Yale University, New Haven, CT 06520, USA}

\author{Joseph N. Burchett}
\affiliation{Department of Astronomy, New Mexico State University, 1320 Frenger Mall,
Las Cruces, NM 88003}

\author{Mohammadreza Ayromlou}
\affiliation{Argelander-Institut f\"ur Astronomie, Auf dem H\"ugel 71, D-53121 Bonn, Germany}

\begin{abstract}
Accurately inferring dark matter halo properties like the total halo mass ($M_{halo}$) and gas fractions ($f_{gas}$) remains particularly challenging at group scales, where low member counts, shallow potential wells, and AGN feedback-driven baryon expulsion introduce significant observational scatter. As large-scale surveys begin to provide unprecedented multi-wavelength data, there is a pressing need for methods that can jointly leverage diverse observables to overcome these uncertainties. We introduce the Deep Learning Halo Definer (DLHD), a multimodal deep learning framework that simultaneously processes galaxy catalogues and X-ray imaging through the combination of a Deep Sets and a Convolutional Neural Network (CNN) to improve $M_{\rm halo}$ and $f_{\rm gas}$ estimation for galaxy groups and clusters. Using mock datasets derived from the IllustrisTNG300 hydrodynamic simulation, we demonstrate that the DLHD outperforms each of its component networks individually, achieving RMSE improvements in $M_{\rm halo}$ of $1.9\times$ over Deep Sets and $1.3\times$ over the CNN alone. For the gas fraction enclosed within $R_{200c}$, DLHD reduces RMSE by $2.0\times$ relative to Deep Sets and $1.1\times$ relative to the CNN alone, with consistent improvements across other apertures. These results highlight a novel ability to leverage multi-band information inaccessible to single-modality methods, positioning DLHD as a promising tool for next-generation survey analyses.

\end{abstract}

\keywords{Neural Networks -- Machine Learning -- Galaxy Groups -- Galaxy Clusters -- X-ray Astronomy -- Intragroup Medium -- Hydrodynamical Simulations}

\section{Introduction}
\label{sec:intro}

Accurate characterization of galaxy groups and clusters is a long-standing observational challenge with direct implications for AGN feedback models, the cosmic baryon budget, and the baryonic systematics of weak lensing surveys. The core difficulty lies in a lack of tools that can jointly exploit the multiple observational channels that current surveys provide.

The most fundamental property of a dark matter halo is its total mass $M_{\rm halo}$. While various techniques exist to measure this, we focus on ``baryonic'' methods, or those that leverage observations of baryons associated with a halo to infer its total mass\footnote{As opposed to Weak Lensing, which measures the gravitational effect of mass using background galaxies.}. One common approach uses luminous galaxies as tracers of the underlying dark-matter-dominated mass: counting the number of galaxies can measure the richness \citep{Rykoff2014,Gonzalez2019}, calculate the line-of-sight velocity dispersions of galaxies \citep{Saro2013,Armitage2018}, or use a phase-space determination \citep{Mamon2013,Li2019_phasespace}, any of which can be used to determine $M_{\rm halo}$. Alternatively, gas properties correlated with $M_{\rm halo}$ (e.g., luminosity and temperature) can be detected by X-ray emission \citep[e.g.,][]{Kravtsov2006,Pratt2009}. 

While these measurements are highly effective for cluster-sized halos, with masses $M_{\rm halo}\ge 10^{14.5}\ {\rm M}_{\odot}$, galaxy group-sized halos ($10^{12.85} \le M_{\rm halo} < 10^{14.5}\ {\rm M}_{\odot}$) present a significant challenge. In X-rays, their luminosity drops precipitously with halo mass, and there is more scatter in their scaling relations \citep[e.g.,][]{akino2022,popesso2025}. The uncertainty of the dynamical mass estimate grows due to lower galaxy counts, higher uncertainty in richness and velocity dispersion measurements \citep[][]{Old2014, Wojtak2018, oppenheimer_simulating_2021,vankempen2025}, and increased difficulty in identifying the true halo center \citep{Berlind2006}.  As large-scale surveys, including {\it eROSITA} in the X-ray \citep{predehl_erosita_2021} and the Dark Energy Survey Instrument (DESI) in the optical \citep{desi2022}, provide unprecedented multi-wavelength data across these mass scales, there is a critical need for a halo parameter estimator that can integrate diverse observables to overcome these inherent 
uncertainties.

Since groups are more abundant than clusters, they have a greater impact on precision cosmology, affecting measurements of cosmic shear \citep{to_2024}, galaxy-galaxy lensing \citep{Meneghetti_2023}, galaxy clusters, redshift space distortions \citep{kaiser_1987,hamilton_1998,Marulli_2017}, and CMB lensing \citep{Madhavacheril_2015,Baxter_2015,Geach_2017,Baxter_2018,Madhavacheril_2020,Sun_2022}. As we enter the era of large galaxy survey science, including ongoing surveys such as SDSS-V \citep{sdss_v_2019}, DESI \citep{desi_2014}, and Euclid \citep{euclid_2019}, alongside X-ray missions like \textit{Chandra} and \textit{XMM-Newton}, the need for tools that can improve the precision of group-scale halo parameter estimators has become urgent.

However, even with these observational surveys, X-ray emission drops rapidly for lower $M_{\rm halo}$ \citep[][]{McCarthy2010,Lovisari2021}, leading to scaling relations that are less constrained and susceptible to higher uncertainty \citep[][]{Babul2002,Osmond2004}. Baryon depletion in groups is widely attributed to AGN feedback, which is thought to expel and redistribute gas, decoupling baryons from dark matter. Because groups have shallower potential wells than massive clusters, AGN feedback is more destructive and drives stronger non-thermal pressure support, biasing hydrostatic X-ray mass estimates \citep{eckert2021}, in contrast to massive clusters where the bias is driven primarily by non-thermal pressure from gas motions associated with dark matter halo assembly \citep[e.g.,][]{Nagai2007,Lau2009,Lau2013,Nelson2014b,Nelson2014a}. AGN feedback also imprints a dynamical noise-like signature in the spatial and velocity distribution of the galaxy population. Importantly, these are not independent noise sources, as the same AGN activity that expels gas also drives non-thermal pressure support that biases hydrostatic X-ray mass estimates, while simultaneously imprinting signatures in the spatial and velocity distributions of the galaxy population that a galaxy-only method cannot disentangle from dynamical noise. Hence, disentangling the feedback signal from noise requires more than a galaxy-only method, thus motivating our unique framework.

Machine Learning (ML) provides a promising path forward. Early ML efforts often focused on single datasets, typically using mock datasets developed from cosmological hydrodynamical simulations for training. For instance, Convolutional Neural Networks (CNNs) have been used to infer halo masses from X-ray imaging \citep{Ntampaka2019}, galaxy catalogue data \citet{ho_robust_2019}, and combined \texttt{CAMELS}-simulated X-ray and HI maps \citep{gluck_2024}. Subsequent tools have expanded these applications to X-ray images, \citep{Sadikov2025}, profiles \citep{Iqbal2025}, and discretized galaxy catalogue data \citep{Wang2020,Ma2025}.  

Modern deep learning architectures, including CNNs, allow for the merging of latent vectors within hidden layers, facilitating ``multimodal" networks. \citet{ho_2023} demonstrated this by combining a 2-D X-ray image CNN with a 3-D dynamical phase space cube to improve cluster mass predictions. Such integration is crucial for maximizing the utility of modern surveys, which provide a wealth of data, including spectroscopic redshifts, star formation rates, and stellar masses. Because galaxy survey catalogue data are arranged as discretized lists of properties, it is ideal for ``point-cloud analysis,'' where each galaxy and its properties are treated as a point in an unordered catalogue.  ML algorithms excel at modeling these clouds and, in cluster applications, have already reduced systematic uncertainty by more than a factor of two \citep{ntampaka2016dynamical, ho_robust_2019}. 

No prior framework has jointly trained a permutation-invariant galaxy group catalogue network and an X-ray imaging CNN within a single end-to-end architecture. In this paper, we introduce the Deep Learning Halo Definer (DLHD), which combines the Deep Sets algorithm \citep{zaheer_deep_2017}, designed to treat unstructured, permutation-invariant galaxy catalogue datasets, with a CNN for X-ray image analysis. By uniting these ``expert" component networks, we achieve a more accurate estimate of $M_{\rm halo}$ and empirically test whether this complementarity yields a meaningful improvement in joint inference over single-modality approaches.

The paper is organized as follows. In \S\ref{sec:methods}, we discuss the datasets that we generate from simulations (\S\ref{sec:mockdatasets}), introduce the neural network architectures, culminating in DLHD (\S\ref{sec:architecture}), and detail the implementations of DLHD and its component networks.  The results are presented in \S\ref{sec:results}, divided into findings for $M_{halo}$ inference by Deep Sets (\S\ref{sec:ds_res}), the CNN from \citet{adams_2026} (\S\ref{sec:cnn_res}), and the DLHD network (\S\ref{sec:comb_res}). We then compare DLHD and its component networks in $f_{gas}$ inference in \S\ref{sec:fgas}. This is followed by a general discussion of our results in \S\ref{sec:discussion}, including a comparison to previous studies (\S\ref{sec:prevresultcomp}) and a discussion of the advantages of multimodality in this type of inference (\S\ref{sec:DLHDperformance}). We conclude with our key results and a discussion of future directions in \S\ref{sec:conclusion}.

\section{Methods} \label{sec:methods}

\subsection{Generation of Mock Datasets} \label{sec:mockdatasets}

We use IllustrisTNG300 (hereafter, TNG300) cosmological hydrodynamic simulation \citep{nelson_first_2018,pillepich_first_2018,marianacci_2018,springel_2018, naiman_2018}, which is a ($302.6$ Mpc)$^3$ volume initially consisting of dark matter (DM) and gas fluid elements at $z=127$ distributed according to the \citet{planck_cosmo_2016} cosmological parameters.  The $2500^3$ DM particles have a mass of $5.9\times 10^7\;\msolar$ and the initial $2500^3$ fluid elements have $1.1\times10^7\;\msolar$.  

The simulation is evolved to $z=0$ with the IllustrisTNG prescriptions for gas cooling, star formation, metal enrichment, stellar feedback, and black hole formation, assembly, and feedback as detailed in \citet{pillepich_simulating_2018} and \citet{weinberger_17}.  We select halos that host galaxy groups from snapshot 99 at $z=0.0$.  Throughout this work, we adopt $M_{500c}$ as our halo mass definition, the mass enclosed within the radius at which the mean interior density is 500 times the critical density of the Universe. Our sample includes 3477 unique halos with mass $10^{12.85}\;\msolar\leq M_{500c} < 10^{14.5}\;\msolar$. The properties we infer are listed in Table~\ref{table:infer_parameters}.  

\begin{deluxetable*}{llcc}
\tablecaption{Properties used by the deep learning models. The \textit{halo properties} are the quantities being inferred, where $f_{\rm gas}$ is the total diffuse gas mass divided by the total mass within various radii (0.5, 0.7, 1, and 2)$\times R_{\rm 500c}$ compiled by \citet{Ayromlou2023}. The \textit{galaxy input properties} are the training data from TNG300, where $R$ is the radial distance measured from the center, and velocity refers to the speed relative to the halo's center.\label{table:infer_parameters}\label{table:gal_parameters}}
\tablewidth{0pt}
\tablehead{\colhead{} & \colhead{\textbf{Property}} & \colhead{\textbf{Range}} & \colhead{\textbf{Median}}}
\startdata
\multirow{5}{6.5cm}{\textbf{Halo properties (inferred)}}
  & $M_{500c}$ [log($\msolar$)] & 12.85--14.48 & 13.15 \\
  & $f_{\rm gas}(0.5R_{200c})$  & 0.008--0.153 & 0.045 \\
  & $f_{\rm gas}(0.7R_{200c})$  & 0.012--0.155 & 0.061 \\
  & $f_{\rm gas}(R_{200c})$     & 0.020--0.160 & 0.081 \\
  & $f_{\rm gas}(2R_{200c})$    & 0.048--0.157 & 0.118 \\
\hline
\multirow{4}{6.5cm}{\textbf{Galaxy input properties (TNG300)}}
  & Mass [log($M_\odot$)] & 9.0--12.7 & 9.7 \\
  & SFR [$M_\odot$/yr]    & 0--45 & 0 \\
  & $R$ [kpc]             & 0--2100 & 701 \\
  & Velocity [$\kms$]       & -$2500$ -- $+2500$ & 0 \\
\enddata
\end{deluxetable*}

\subsubsection{Galaxy Catalogues} \label{sec:GalaxyCat}

To generate galaxy catalogues, we use the \texttt{SUBFIND} group catalogue \citep{springel_subfind_2001} downloaded from the IllustrisTNG website containing the TNG300 dataset\footnote{\href{https://www.tng-project.org/data/downloads/TNG300-1/}{https://www.tng-project.org/data/downloads/TNG300-1/}} for the $z=0.0$ snapshot.  We create 3 mock catalogues along the principal axes ($x$, $y$, and $z$).  We include all galaxies within $\pm 2500\;\kms$ of the halo's central velocity, and within a square field 3000 kpc across.  We catalogue stellar mass ($M_\star$) and star formation rate (SFR) for all galaxies $M_\star\geq 10^9\;\msolar$.  A summary of the galaxy input parameters is listed in Table~\ref{table:gal_parameters}.

\subsubsection{X-ray Observations}

\begin{figure*}
  \centering
  \includegraphics[scale=0.35]{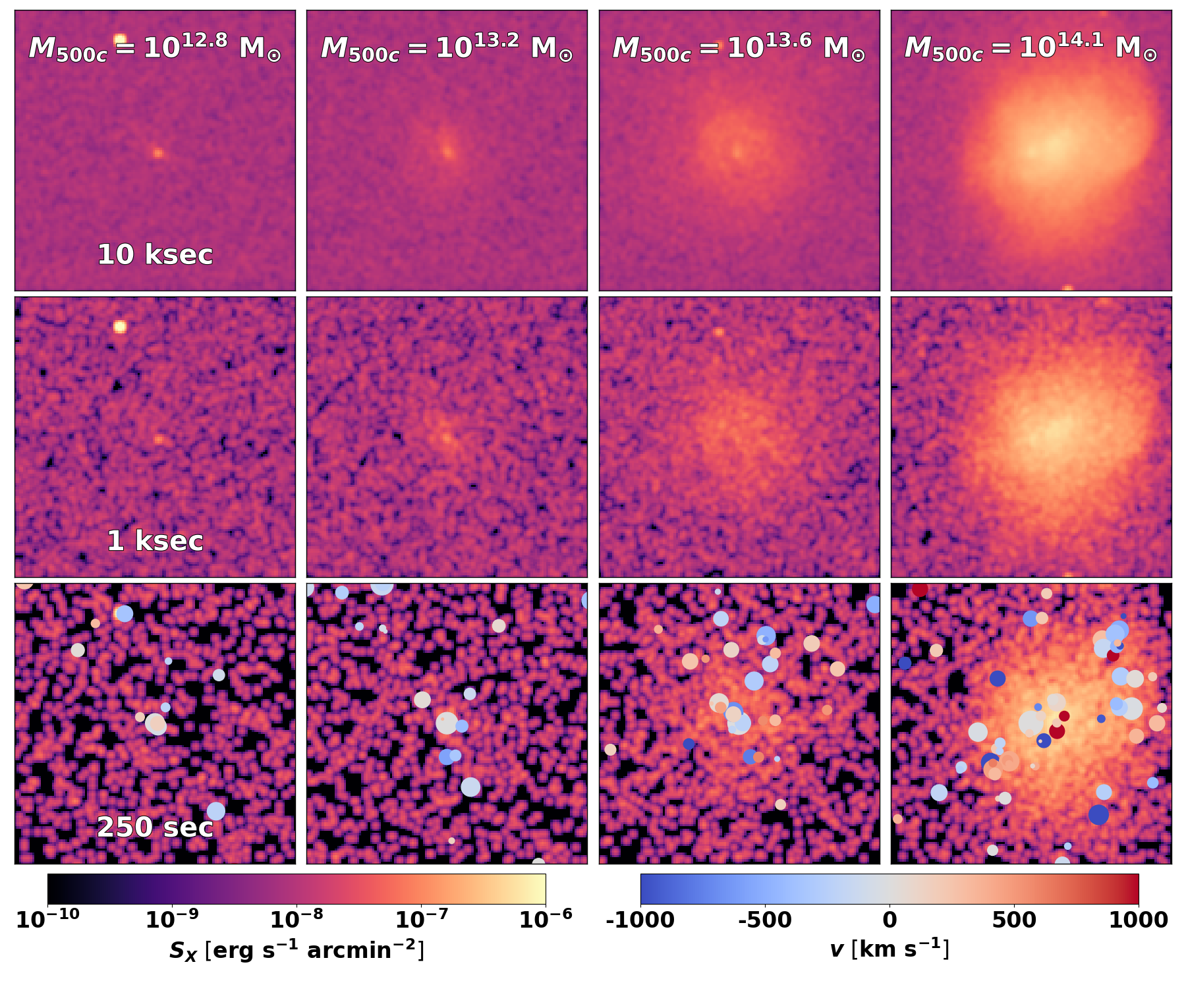}
  \caption{X-ray mocks for four TNG300 groups with mass increasing from left to right. Exposure length decreases from top to bottom.  We show the galaxy catalogue data as positions in the bottom panel, colored by velocity, with the point size scaled by stellar mass.  }
  \label{fig:mockdata}
\end{figure*}

We use the pyXSIM software package \texttt{pyXSIM} \citep{zuhone_pyxsim_2016} to generate a mock photon list for every TNG300 halo. Then, we use the \texttt{SOXS} package \citep{zuhone_soxs_2023} to mock soft X-ray observations from 0.5-2.0 keV. We approximate an X-ray telescope with a collecting area of 2000~cm$^2$, and a point spread function (PSF) similar to {\it eROSITA} \citep{predehl_erosita_2021} and {\it XMM Newton} \citep{xmm_newton_2022}. The field of view spans 32 arcminutes using 128$\times$128 $15''$ pixels, which approximates the collecting area of these two telescopes.  

We assume a redshift of $z=0.035$, which is a distance of approximately 150 Mpc.  We simulate exposures of three lengths: 250 seconds, 1 kilosecond (ksec), and 10 ksec.  The first exposure length is comparable to the {\it eROSITA} eRASS:1 observations, while the second is meant to simulate the 2-year eRASS:4 survey.  The last one simulates a deeper survey achievable with {\it XMM} \citep[e.g.,][]{Eckert2024} or the eRASS:4 survey at the ecliptic poles.

We add instrumental background and emission from the Milky Way Galactic component to our group mocks.  We mock a background-only field, and subtract the two to create a background-subtracted ``source'' mock.  We require positive values for CNN; therefore, we set all pixels with negative or zero values to 0.5, while other pixels are positive integer counts. Altogether, we generate four different exposure lengths for three axes across 3477 halos.  

We show mock images of the source photons from four groups of increasing masses, from left to right in Figure~\ref{fig:mockdata}. The field of view is 1380~kpc. Exposure times decrease from 10~ksec in the first row, to 1~ksec in the second, and 250 seconds in the third. Morphological features, including shock fronts associated with a merger in the $10^{14.1}\;\msolar$ halo, are more clearly observable in longer exposures. In the third row (250-second exposure), we overplot the galaxy positions from our galaxy catalogue, colored by the offset velocity from the group center.  

\subsection{Deep Learning Halo Definer}\label{sec:architecture}

\begin{figure*}[t]
    \gridline{
        \fig{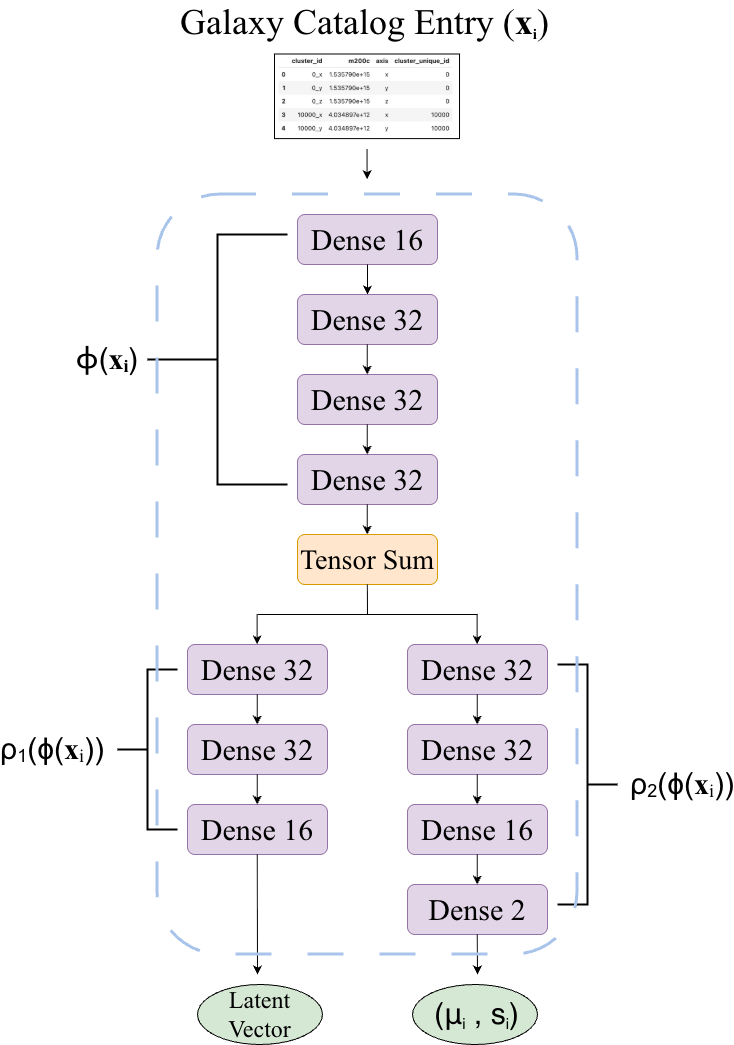}{0.38\textwidth}{(a) Deep Sets}
        \fig{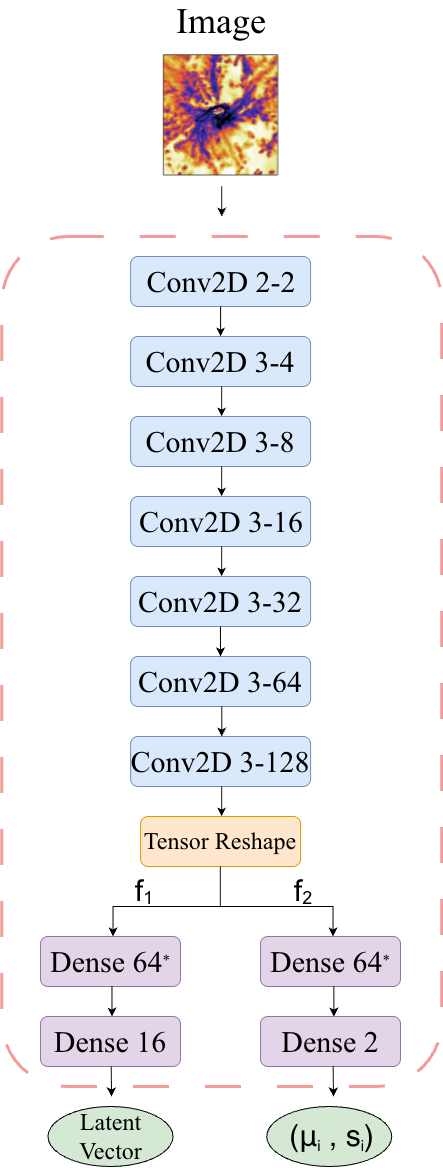}{0.25\textwidth}{(b) CNN}
        \fig{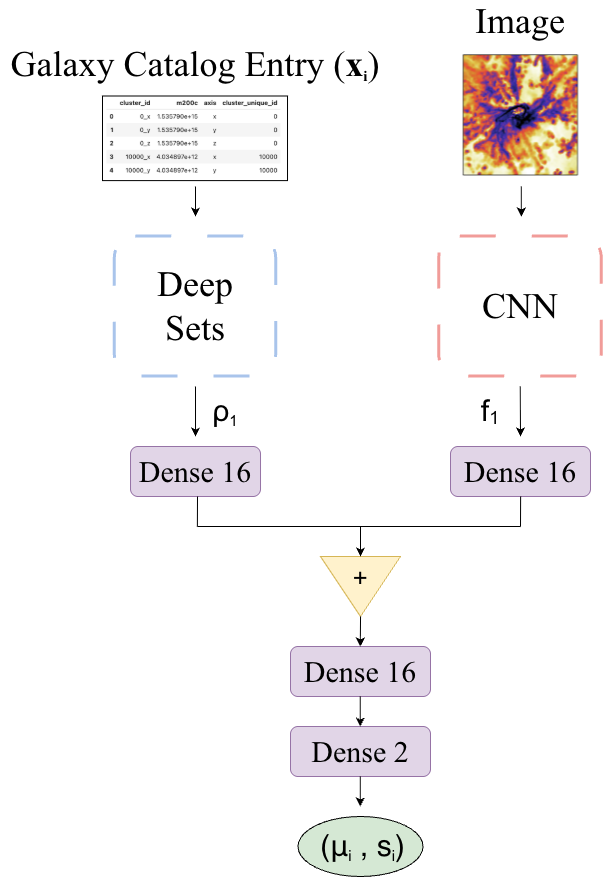}{0.32\textwidth}{(c) DLHD \label{fig:dlhd_ind}}
    }
    \caption{Network architectures used in this work. a) The Deep Sets architecture uses the $\phi$ operator on galaxy data, and the $\rho$ operator processes combined cluster objects with two forward passes: $\rho_1$ used in full DLHD architecture training and $\rho_2$ used in independent component network training. b) The Convolutional Neural Network (see \citet{adams_2026}) operates on X-ray images with two forward passes: $f_1$ used in full DLHD architecture training and $f_2$ used in independent component network training. The linear layers with an asterisk denote layers with dropout, which we set to 0.2. c) The DLHD combines these two component networks through an element-wise multiplication of the resultant latent vectors of forward passes $\rho_1$ and $f_1$, denoted by the `` + '' symbol. This combined vector is then reduced through two dense linear layers into a parameter prediction and a logarithmic uncertainty.}
    \label{fig:architectures}
\end{figure*}

The Deep Learning Halo Definer (DLHD) is a multimodal deep learning architecture. Multimodal machine learning involves constructing a neural network architecture that can process and extract information from multiple data modalities, often necessitating the integration of several component networks into a single architecture \citep{baltrusaitis_2017}. This section details the component networks in the DLHD and the methods used to train and test them. Figure~\ref {fig:architectures} shows a diagram of each component network and the combined DLHD architecture.

\subsubsection{Deep Sets}\label{sec:ds}

The first component of the DLHD is the Deep Sets network. Deep Sets is a permutation invariant neural network architecture designed for finite sets of information \citep{zaheer_deep_2017}. In the case of the DLHD, Deep Sets operates on galaxy catalogues, per \S\ref{sec:GalaxyCat}, and references a corresponding galaxy cluster catalogue for true parameter values while training.

The Deep Sets network consists of two sequential neural transformations: $\phi$ and $\rho$. Transformation $\phi$ maps the properties of a single galaxy to a summary vector, and transformation $\rho$ maps the properties of an entire cluster to a parameter estimate and uncertainty. In practice, these two operations take the form of groups of fully connected (dense) linear layers. 

A single batch $X \in \mathbb{R}^{N_{gal}\times d}$ contains $N_{gal}$ total galaxies, each with $d$ features. These galaxies are grouped into $N_c$ clusters, where the cluster $c$ contains galaxies $X_c$. We used a batch size of $N_c =64$ clusters containing $N_{gal} \in [997,2377]$ total galaxies, each with $d=5$ features: positions $x$ and $y$, velocity, stellar mass ($M_{\rm \star}$), and star formation rate (SFR).

A forward pass $f(X)$ involves first inputting $X$ into $\phi$, then splitting the output by cluster and summing the latent data of each cluster. The resulting vectors are concatenated into a single latent vector passed into $\rho$,
 \begin{equation}
    f(X) = \rho \left( 
    \bigoplus^{N_c} \left( \sum_{x_i \in X_c}\phi(x_i) \right)
    \right),
\end{equation}
in which $\bigoplus$ represents the concatenation of the summed cluster vectors. The output of a batch of $N_c$ clusters is a set of latent vectors $V$, where each vector $v_i \in \mathbb{R}^{N_c \times 2}$.

\subsubsection{CNN}\label{sec:cnn}
The second component of the DLHD is a convolutional neural network (CNN) designed to reduce a batch of $N$ 128$\times$128 pixel images to a set of latent vectors $V \in \mathbb{R}^{N \times 2}$. Our CNN passes the data through seven sets of 2D convolutional layers. The convolutional output is reshaped into a single latent vector, which then passes through two fully connected linear layers. For a more extensive description, see the companion paper \citep{adams_2026}.

\subsubsection{DLHD Architecture}\label{sec:dlhd_arch}

The DLHD combines the aforementioned Deep Sets and CNN into a single architecture. Each component network contains a standard forward pass that returns latent vectors $v_i \in \mathbb{R}^2$, along with an additional forward pass function that returns $v_i \in \mathbb{R}^{16}$. A forward pass through the DLHD utilizes the latter of these component forward passes. The CNN and Deep Sets output vectors are then multiplied element-wise to form a combined information latent vector \citep{zhu_deeponet_2023}, which is passed through two fully connected linear layers and reduced to a two-dimensional vector.

\subsection{Training of Networks}\label{sec:training}

The DLHD training process is twofold: a training loop for each component network, followed by a loop for the full architecture. Both training loops follow the same general procedure and use the same loss functions. This staged design addresses a well-documented challenge in multimodal learning, where modalities that converge at different rates can cause the faster-converging branch to dominate the joint loss and suppress gradient updates to the slower branch, reducing the risk that fusion collapses to a single-modality solution \citep{abdelsalam_ismail_improving_2020}.

As discussed in \S~\ref{sec:architecture}, both individual component networks and the full DLHD architecture output a two-dimensional vector. Through the construction of the following loss functions, these two dimensions represent a parameter estimate $\mu$ and a log-standard deviation $s$. Thus, each latent vector is written as $v_i = (\mu_i,s_i)$. For a batch of size $N$, we denote the set of true parameter values by $\textbf{y}= \{y_i\}_{i=1}^{N}$, the set of predicted parameter values by $\boldsymbol{\mu}= \{\mu_i\}_{i=1}^{N}$, and the set of logarithmic standard deviations by $\boldsymbol{s} = \{s_i\}_{i=1}^{N}$.

During \textit{pre-training}, we use a mean square-error (MSE) loss, $\mathcal{L}_{\text{MSE}}$, that is a function of $\textbf{y}$ and $\boldsymbol{\mu}$, given by
\begin{equation}\label{eq:MSE}
    \mathcal{L}_{\text{MSE}}(\textbf{y},\boldsymbol{\mu}) = \sum_{i}(y_i - \mu_i)^2,
\end{equation}
where a smaller MSE indicates a more accurate network performance.  MSE pretraining drives each branch toward accurate point estimates of $M_{500c}$ and $f_{\rm gas}$ before joint training begins, stabilizing the network weights prior to the more demanding training phase.

Then, during \textit{training}, we use the negative log-likelihood (NLL) loss function, $\mathcal{L}_{\text{NLL}}$, which is a function of $\textbf{y}$, $\boldsymbol{\mu}$, and $\boldsymbol{s}$. The purpose of $\mathcal{L}_{\text{NLL}}$ is to construct a Gaussian distribution with the network outputs $\boldsymbol{\mu}$ and $\boldsymbol{s}$ with which we can evaluate the likelihood of the true parameter values $\textbf{y}$ under this distribution, $p(\textbf{y}|\boldsymbol{\mu},\boldsymbol{s})$. As a function of standard deviations $\boldsymbol{\sigma}=\{ \sigma_i \} = \{ \exp(s_i)\}$, the NLL is given by

\begin{equation}
    \label{eq:gaussian}p(\textbf{y}|\boldsymbol{\mu},\boldsymbol{\sigma}) = \prod_{i} \frac{1}{\sqrt{2\pi \sigma_i ^2}} \exp \left(- \frac{1}{2} \left(\frac{y_i - \mu_i}{\sigma_i}\right)^2\right).
\end{equation}

Taking the negative natural-logarithm of Equation~\ref{eq:gaussian} gives $\mathcal{L}_{\text{NLL}}(\textbf{y},\boldsymbol{\mu},\boldsymbol{\sigma})$,
\begin{equation}
     \mathcal{L}_{\text{NLL}}(\textbf{y},\boldsymbol{\mu},\boldsymbol{\sigma}) \propto  -\sum_i \left(-\log(\sigma_i) - \frac{1}{2}\left(\frac{y_i - \mu_i}{\sigma_i} \right)^2 \right).
\label{eq:NLL}
\end{equation} 
Note that the $2\pi$ factor is dropped because it does not change the location of the minimum of $\mathcal{L}_{NLL}$. The NLL loss produces per-object uncertainty estimates alongside each prediction; we validate the calibration of these uncertainties on simulation mocks, and recalibration will be required before deployment on real survey data, where the noise model differs from the training distribution.

In training, our goal is to maximize the likelihood $p(\textbf{y}|\boldsymbol{\mu},\boldsymbol{\sigma})$, or in other words, to minimize the negative log-likelihood $\mathcal{L}_{\text{NLL}}$. Hence, a more negative value of $\mathcal{L}_{\text{NLL}}$ is interpreted as a more accurate model. This loss function can also be expressed in terms of the log standard deviation $\textbf{s}=\{s_i\}_{i=1}^{N} = \log(\boldsymbol{\sigma})$,
\begin{equation}
     \mathcal{L}_{\text{NLL}}(\textbf{y},\boldsymbol{\mu},\boldsymbol{s}) \propto  -\sum_i \left(-s_i - \frac{1}{2}\left(\frac{y_i - \mu_i}{\exp(s_i) + \epsilon} \right)^2\right),\label{eq:std_NLL}
\end{equation} 
where $\epsilon=10^{-5}$ is a softening parameter that prevents the loss function from diverging when the predicted logarithmic uncertainty is very small. 
We use the form of $\mathcal{L}_{\text{NLL}}$ in Equation~\ref{eq:std_NLL} in our computations due to numerical stability, and note that the standard deviations $\sigma_i$ are recovered via $\sigma_i = \exp(s_i)$ for the error bars in the truth-prediction plots shown in Figures~\ref{fig:DS_m500_vary}-\ref{fig:DLHD_m500c_varyx} and Figure~\ref{fig:comp_fgas}. It is through this loss function that the two-dimensional outputs of each neural network is interpreted as a parameter estimate and logarithmic uncertainty.

\subsubsection{Training}\label{sec:algorithms}
A complete DLHD training loop can be divided into three phases: 1) pretraining the Deep Sets component network, 2) pretraining the CNN component network, and 3) training the full DLHD architecture. The dataset was split into 80\%-20\% for training and testing respectively. We implement early stopping in all training phases, updating model weights only when the validation loss improves by at least $10^{-3}$ over the previous epoch, retaining the best-performing model at the end of each phase, and terminating the training run if this condition is not satisfied after 30 epochs. We chose this margin to be small enough to encourage refinement while still large enough to prevent overfitting. 

All phases also utilized \texttt{Optuna} \citep{optuna_2019} to perform a hyperparameter search on the learning rate and the weight decay, using a range of $[10^{-5},\, 10^{-3}]$ for each parameter. Each trial ran for 30 epochs, selecting the configuration that minimized the root mean square error on the testing set at the end of each trial. The optimal learning rates and weight decay values across all network configurations ranged from $1.85 \times 10^{-4}$ to $9.99 \times 10^{-4}$ and $1.02 \times 10^{-5}$ to $9.99 \times 10^{-4}$, respectively, with the full trial results available in the accompanying GitHub repository \href{https://github.com/CalebBOgle/DLHD}{https://github.com/CalebBOgle/DLHD}.

Each component network is pretrained for 100 epochs using a simple MSE loss to stabilize the network's weights and biases \citep{skafte2019reliable}, followed by 100 training epochs using the NLL loss (Eq.~\ref{eq:NLL}). Full DLHD training begins with these component pretraining steps, followed by end-to-end pretraining and training, for a total of 600 epochs. This takes $\mathcal{O}(\text{2 hrs})$ on a single CPU, unless the early stopping criteria are met.

\subsubsection{Performance Metrics}\label{sec:metrics}

After training, we quantify the network's ability to infer a parameter from the unseen test set using the root mean square error (RMSE). For a batch of $N$ parameter estimates, this is given by the standard deviation of the difference between the values of the predicted ($\mu_i$) and true ($y_i$) parameters, 
\begin{equation}
    {\rm RMSE}  = \sqrt{\frac{1}{N}\sum_{i=1}^{N}(\mu_i-y_i)^2}\label{eq:scatter},
\end{equation}
where a lower RMSE indicates a more accurate and thus better-performing network. 

We also use the coefficient of determination $R^2$ to quantify the performance of the network, defined as
\begin{equation}
    R^2 = 1-\frac{\sum_{i=1}^{N_c}(y_i - \mu_i)^2}{\sum_{i=1}^{N_c}(y_i - \bar{y_i})^2},
\end{equation}  
in which $\bar{y}$ is the mean value of the estimated parameter. With this statistic, values closer to 1.0 indicate more accurate estimates. 

Finally, we used the reduced chi-squared statistic to quantify the reliability of the network-predicted errors $\sigma_i = \exp(\text{s}_i)$,

\begin{equation}
    \label{eq:reduced chi squared}
    \chi^2 = \frac{1}{N} \sum_{i=1}^{N} \left(\frac{y_i - \mu_i}{\sigma_i} \right)^2.
\end{equation}
With this, a value close to one is interpreted as properly quantified errors, and hence a well calibrated network, whereas values over (under) one indicate underestimated (overestimated) errors.

\section{Results}\label{sec:results}

We first explore the performance of Deep Sets in \S\ref{sec:ds_res} for $M_{500}$, followed by the CNN for the same property in \S\ref{sec:cnn_res}. We then demonstrate that the DLHD network performs best for halo mass inference compared to its component networks in \S\ref{sec:comb_res}. We also show the results from gas fraction estimation by these networks in \S\ref{sec:fgas}. 

\subsection{Deep Sets Network}\label{sec:ds_res}

\begin{deluxetable}{lccc}
\tablecaption{Deep Sets results inferring $M_{500c}$, with the full training set results shown in the first row and data quality cuts shown in subsequent rows.\label{table:DS_m500_vary}}
\tablewidth{0pt}
\tablehead{\colhead{\textbf{Galaxy Input Change}} & \colhead{\textbf{RMSE}} & \colhead{$\mathbf{R^2}$} & \colhead{$\mathbf{\chi^2}$}}
\startdata
-- & 0.104 & 0.906 & 1.209 \\
\hline
$M_{\rm gal} > 10^{9.5} M_{\odot}$  & 0.104 & 0.902 & 1.074 \\
$M_{\rm gal} > 10^{10} M_{\odot}$   & 0.105 & 0.906 & 1.186 \\
\hline
$\sigma_c = 50 ~\text{kpc}$   & 0.169 & 0.723 & 1.174 \\
$\sigma_c = 100~\text{ kpc}$  & 0.162 & 0.750 & 0.885 \\
$\sigma_c = 250~\text{ kpc}$  & 0.180 & 0.728 & 0.938 \\
$\sigma_c = 500~\text{ kpc}$  & 0.190 & 0.697 & 1.210 \\
\hline
$\sigma_v = 30\text{$~\kms$}$   & 0.160 & 0.772 & 1.103 \\
$\sigma_v = 100\text{$~\kms$}$  & 0.170 & 0.731 & 1.369 \\
$\sigma_v = 300\text{$~\kms$}$  & 0.172 & 0.714 & 1.184 \\
$\sigma_v = 1000\text{$~\kms$}$ & 0.178 & 0.710 & 1.014 \\
\enddata
\end{deluxetable}

\begin{figure*}

\includegraphics[width=\textwidth]{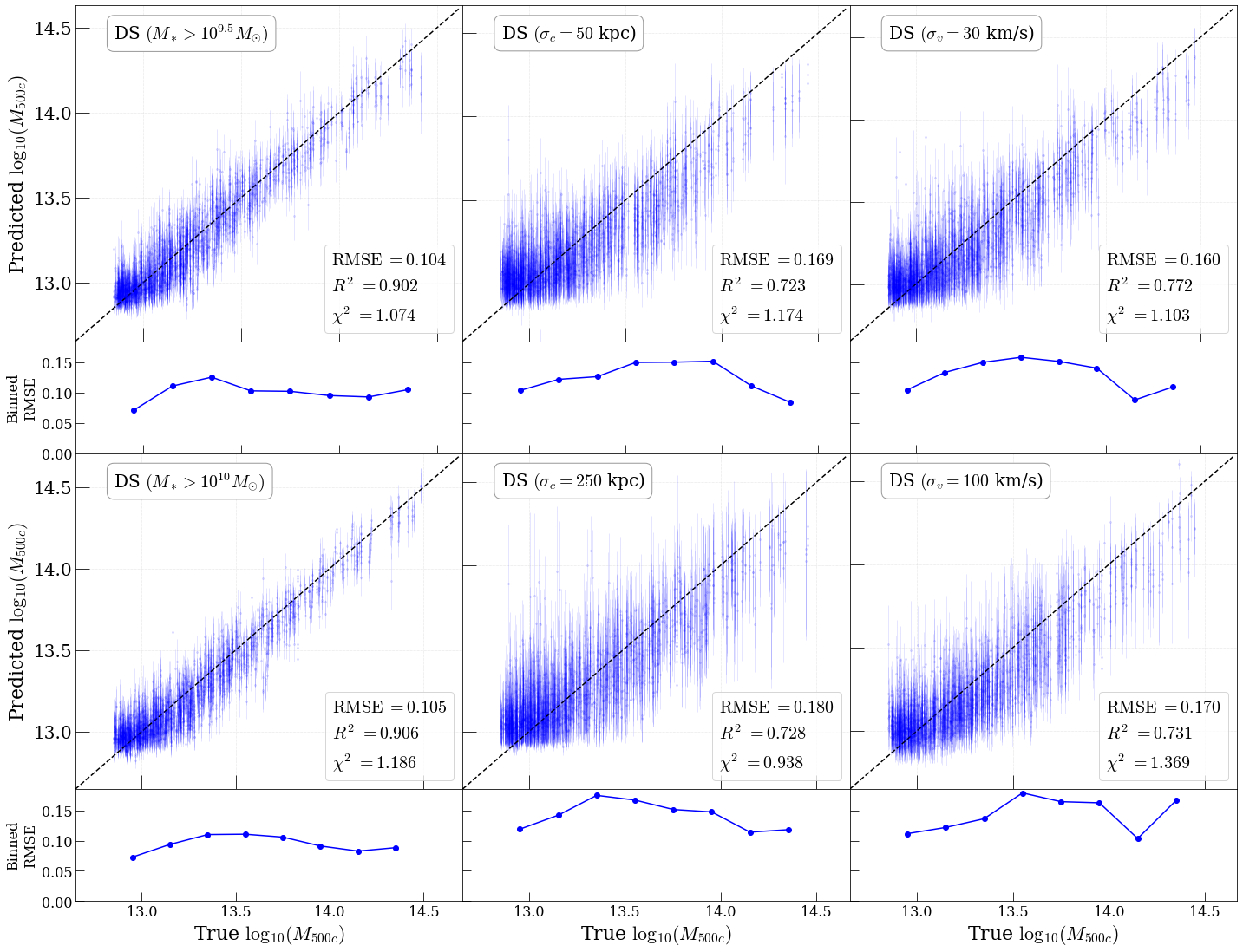}
\caption{Truth-prediction plots (large panels) using the Deep Sets network to infer $M_{500c}$ for various degraded datasets, by column. The small panels below show the corresponding RMSE values across the mass bins. \textit{Left:} Results based on two stellar mass cuts, $10^{9.5}\;\msolar$ and $10^{10}\;\msolar$). \textit{Middle:} Results based on adding mis-centering uncertainty of the central galaxy by $\sigma_c=50\;$kpc, and $\sigma_c=100\;$kpc. \textit{Right:} Results based on adding velocity uncertainty to individual galaxies, of $\sigma_v=30\;\kms$, and $\sigma_v=100\;\kms$.}
\label{fig:DS_m500_vary}
\end{figure*}

We first focus on the results of $M_{500c}$ estimation by the Deep Sets network to determine whether performance degrades when applying various galaxy-mass cuts to the training set. The initial data contains galaxies with $M_{\star}\geq10^9M_{\odot}$ per Table~\ref{table:gal_parameters}. Two additional cuts are applied to these training data: the least aggressive cut removes all galaxies with $M_{\star}<10^{9.5}\ M_{\odot}$, and the most aggressive cut removes all galaxies with $M_{\star}<10^{10}\ M_{\odot}$, corresponding to the removal of approximately 38\% and 66\% of all galaxies, respectively. These removals are meant to mimic a less powerful observational survey that cannot resolve lower-mass galaxies.

The results of this test are shown in the left column of Figure~\ref{fig:DS_m500_vary}. The top panel shows the Deep Sets results (truth-prediction and binned RMSE) when trained on galaxy catalogues with the least aggressive cut, and the lower panel shows the results with the most aggressive cut. Surprisingly, the performance does not worsen significantly when we limit the minimum galaxy mass. This is likely because the performance of Deep Sets is dominated by the most massive galaxy in each cluster, which the trials do not mask out. The RMSE, $R^2$, and $\chi^2$ statistics from the trials on the initial data and the two data cuts are summarized in rows 1-3 of Table~\ref{table:DS_m500_vary}.

\begin{figure*}

\includegraphics[width=\textwidth]{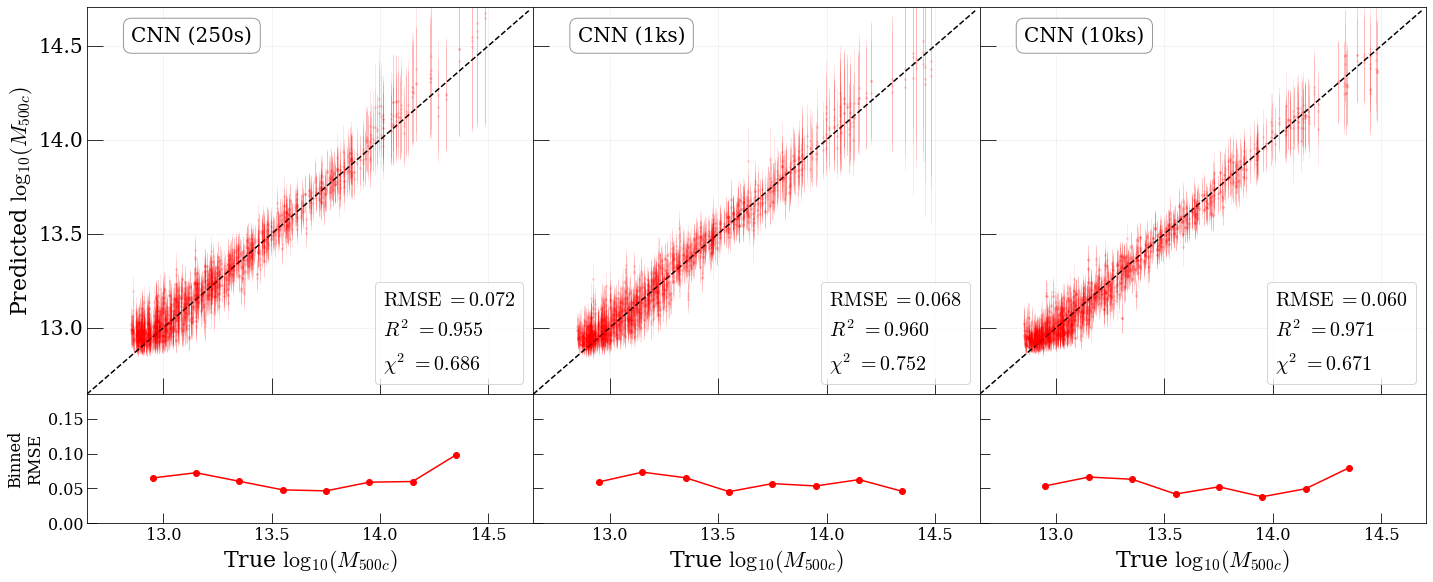}
\caption{Truth-prediction plots for $M_{\rm 500c}$ (large panels) and binned RMSE (small panels) for the CNN, with increasing exposure times from left to right: 250 seconds, 1~ksec, and 10~ksec.}
\label{fig:cnn_m500}
\end{figure*}

Next, we demonstrate the impact of mis-centering galaxies on the $M_{\rm 500c}$ inference to understand how Deep Sets responds to degraded data in which the center of each galaxy cluster is ill-defined. To do this, we generate independent random positional noise values for the $x$ and $y$ directions and shift the positions of every galaxy in a particular cluster by those amounts, repeating this process for every cluster. These noise values are generated by sampling a normal distribution, $\mathcal{N}(0,\sigma_c)$, with $\sigma_c$ being a free parameter used to control the level of uncertainty in the position of the cluster's center.

The middle column of Figure~\ref{fig:DS_m500_vary} shows the results of this mis-centering. The top panel shows a mis-centering characterized by  $\sigma_c= 50$ kpc, while the bottom panel shows a mis-centering characterized by $\sigma_c = 100$ kpc. We also tested higher positional uncertainties characterized by $\sigma_c = 250$ kpc and $\sigma_c = 500$ kpc, and while these results are not plotted, they are reported in Table~\ref{table:DS_m500_vary}. We find that perturbing the assigned center makes it harder for the Deep Sets model to identify the central galaxy. As we saw in the galaxy-mass cut results, the central galaxy has the greatest influence on the inference. Thus, any error in detecting the central galaxy's properties negatively impacts halo-mass inference. This is exacerbated significantly when $\sigma_c$ increases, suggesting that our results are sensitive to mis-centering.  Positional errors in the group center using galaxies can be significant, often with mis-centerings of hundreds of kpc when there are few group members \citep[cf.][]{Berlind2006,stocke2019}. When a central galaxy is clearly identified near the X-ray centroids, it traces the center of mass to $<75\;$kpc, compared to 100-150 kpc with larger offsets \citep{George2012}. Table~\ref {table:DS_m500_vary} summarizes the RMSE, $R^2$, and $\chi^2$ statistics for all four levels of miscentering in rows 4-7.

Lastly, we apply error dispersion for velocities to measure our inference's sensitivity to errors in observed spectroscopic redshift measurements. Unlike positional noise (which was kept constant for each galaxy in the clusters), we add an independent random velocity noise value to each of the galaxies in the clusters, drawn from a normal distribution, $\mathcal{N}(0,\sigma_v)$. We vary the standard deviation of this distribution $\sigma_v$ to measure the impact of reduced spectroscopic performance and even photometric redshifts with a maximum of $\sigma_{v} = 1000\;\kms$. As stated in \S\ref{sec:GalaxyCat}, we include all galaxies within $\pm 2500\;\kms$, so the performance can be affected by contamination from unassociated galaxies.

The right column of Figure~\ref{fig:DS_m500_vary} visualizes the results of including noise in the velocity. The top panel shows the effect of adding noise characterized by $\sigma_v=30~\kms$, and the bottom panel adds noise with $\sigma_v=100~\kms$. We choose the lower noise value to align with the typical velocity errors for current and upcoming surveys, which report errors of the order of $\sigma_{v}\sim 30\;\kms$ \citep[e.g.,][]{desi2022}. While not plotted, we also investigated more extreme uncertainties of $\sigma_v = 300~\kms$ and $\sigma_v = 1000~\kms$. Increasing the noise degrades the precision of $M_{\rm 500c}$ inference across the entire mass range explored here. This suggests that DS uses group dynamics to probe the total system mass, in addition to the constituent galaxy properties. The RMSE, $R^2$, and $\chi^2$ statistics of all four levels of velocity noise are summarized in rows 8-11 of Table~\ref{table:DS_m500_vary}.

\subsection{Convolutional Neural Network} \label{sec:cnn_res}

We now discuss the results of the CNN applied to the simulated X-ray images to compare its predictive power with that of the Deep Sets network when each is used independently, and to obtain a baseline for the DLHD to improve upon.  Figure~\ref{fig:cnn_m500} shows the truth-prediction results for the CNN predicting $M_{\rm 500c}$ with three exposure times: 250~seconds, 1~ksec, and 10~ksec, from left to right. Compared to the best isolated-DS run, the CNN consistently outperforms it in both RMSE and $R^2$ across all exposure lengths. This is somewhat expected, as \citet{ho_2023} showed that when utilizing {\it eROSITA}-like noise, cluster masses were more precisely predicted with map-level information than from galaxy dynamics. 

We also find that the $\chi^2$ values produced from the CNN's predictions are consistently smaller than one, implying that the CNN is overestimating its own uncertainty. This is consistent with the findings of \citet{adams_2026}, in which a more thorough analysis of the same CNN architecture produced similarly small $\chi^2$ values across a number of predicted halo parameters and observational channels. NLL loss functions like Equation~\ref{eq:std_NLL} can encourage inflating uncertainties for predicted values that are far from the corresponding true values \citep{seitzer_pitfalls_2022}, which is likely causing the dramatic error bars for the inaccurate predictions in the high-mass regions of Figure~\ref{fig:cnn_m500}, thus decreasing the overall $\chi^2$ value.

 Finally, we note that while we mock X-ray exposures with background noise and perform background subtraction, we do not use SIXTE software \citep{sixte} as \citet{ho_2023} did, nor do we include non-diffuse X-ray sources such as X-ray AGN and X-ray binary emission.  

\subsection{DLHD Combined Network} \label{sec:comb_res}

\begin{figure*}

\includegraphics[width=\textwidth]{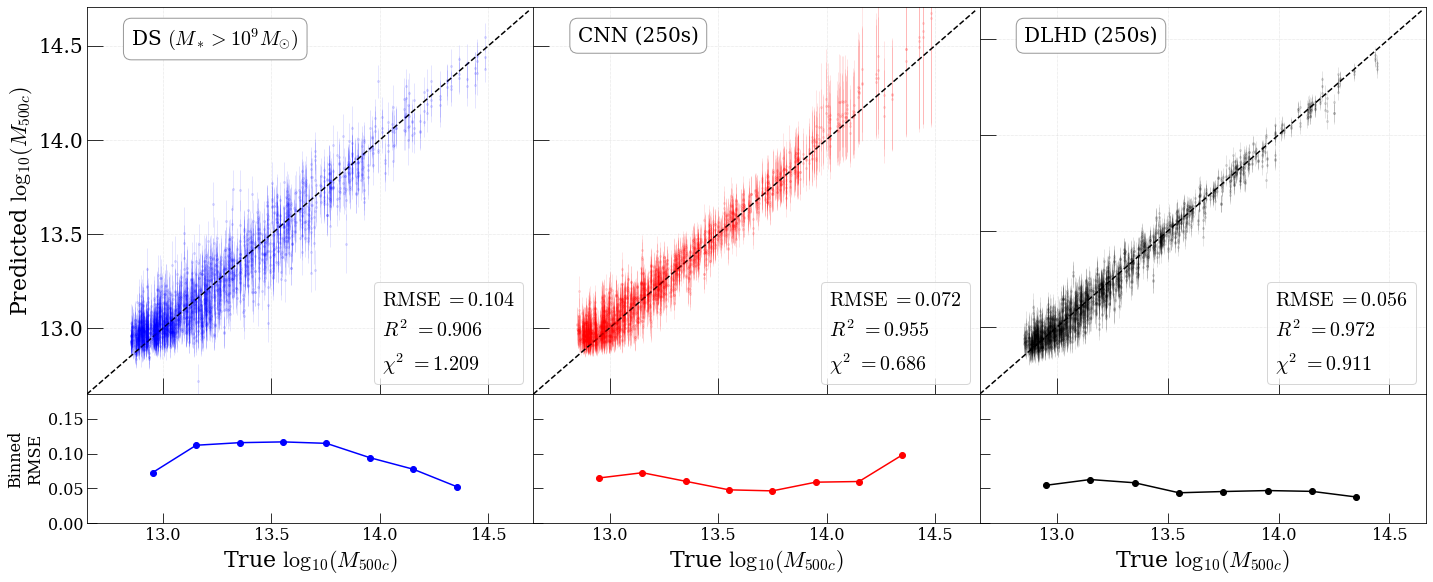}
\caption{From left-to-right, truth-prediction plots and their running binned RMSE in the lower panel for Deep Sets (DS), the CNN, and the combined DLHD network when inferring $M_{500c}$ with idealized galaxy catalogues ($M_\star\geq 10^{9}\;\msolar$, $\sigma_c = 0$ kpc, $\sigma_v= 0\;\kms$) and 250-second X-ray mocks. This demonstrates that the DLHD combined network outperforms the DS galaxy catalogue network and the image-based CNN in terms of accuracy (RMSE,$R^2$) and calibration ($\chi^2$).}
\label{fig:comp_m500}
\end{figure*}

Figure~\ref{fig:comp_m500} shows the main comparison among Deep Sets, the CNN (250-second exposures), and the combined DLHD architecture (250-second exposures). We note that the RMSE statistic greatly improves by combining catalogue data (0.102~dex) and image data (0.072~dex) to achieve 0.056~dex via DLHD. The $R^2$ and $\chi^2$ statistics also show improvement with the DLHD. This shows that adding X-ray imaging even at the level of the 6-month {\it eROSITA} All-Sky Survey 1 (eRASS:1) can improve halo mass inference by a factor of $1.8\times$ upon catalogue-only inference.  
At 0.056~dex (approximately 14\% fractional mass uncertainty) on idealised mocks, the DLHD reaches a precision level that would be scientifically useful for group catalogue science, where current X-ray scaling relation mass estimates for individual groups typically carry uncertainties of 20-40\%. 
 

We also find that this result is an improvement upon a simpler utilization of the Deep Sets and CNN architectures. We combined the predictions for $M_{500c}$ from an independent Deep Sets network and CNN into one larger set, on which we computed the RMSE, $R^2$, and $\chi^2$ to compare with the full DLHD architecture. This naive combination yielded an RMSE of 0.090~dex, which is 1.6 $\times$ larger than that of the DLHD (0.056~dex), thus substantiating the novelty of the method of component network combination in the DLHD architecture. We refer to the GitHub repository for further details. 

These results demonstrate the main purpose of developing the DLHD, and strongly suggest that the residual errors of the Deep Sets network and the CNN are correlated, and hence the full DLHD architecture can capture non-linear information that neither component network alone can model.  If the residual errors of Deep Sets and the CNN are not correlated, the DLHD scatter would not be better than the `best' model among the two. Instead, our results suggest that there are correlated residuals in the mass scatter of one component network that can be modeled by the other. We further analyze how the correlations in the component galaxy catalogue and X-ray imaging datasets can combine to provide a better mass proxy in \S\ref{sec:DLHDperformance}. 
   
\begin{figure*}
\begin{center}
\includegraphics[width=0.67\textwidth]{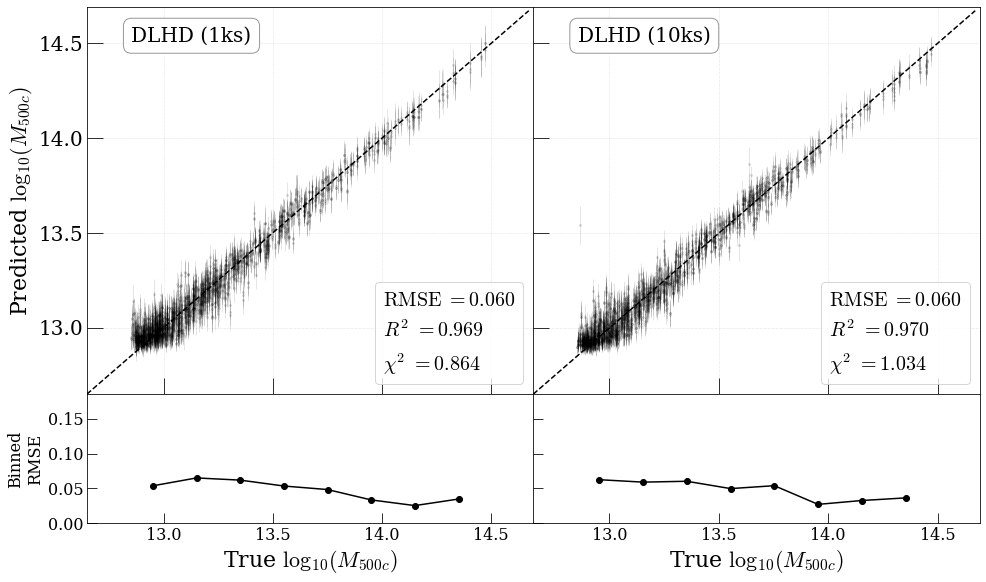}
\end{center}
\caption{Truth-prediction plots (top panels) and binned RMSE (lower panels) using the DLHD architecture when predicting $M_{500c}$ from 1.0~ksec and 10~ksec exposures, which can be compared to the shorter 250-second exposures in the right panel of Fig. \ref{fig:comp_m500}.}
\label{fig:DLHD_m500c_varyx}
\end{figure*}

In addition to the 250-second exposures, we also train the DLHD to infer $M_{500c}$ on 1~ksec and 10~ksec X-ray exposures. The results from these trials are shown in Figure~\ref{fig:DLHD_m500c_varyx} and can be compared to the CNN-only trials shown in Figure~\ref{fig:cnn_m500}. We find that the DLHD maintains its accuracy across exposure lengths, giving an RMSE of 0.060~dex for both 1~ksec and 10~ksec exposures, while the CNN reaches this accuracy only at 10~ksec. The DLHD also appears more calibrated than the CNN across exposure lengths, with its $\chi^2$ values being much closer to one than the CNN. 

The statistical results for $M_{500c}$ estimation of Deep Sets, the CNN, and the DLHD across all mentioned exposure lengths are summarized in the upper portion of Table~\ref{table:DS_CNN_DLHD}.

\setlength{\tabcolsep}{10pt}
\begin{deluxetable*}{lccccccccc}
\tablecaption{RMSE (in dex), $R^2$, and $\chi^2$ values for Deep Sets (DS), the CNN, and the DLHD across the different estimated halo parameters. For the $M_{500c}$ results (rows 1-3), the results from Deep Sets are repeated since it does not train on images, and hence are independent from exposure lengths. For the $\fgas$ results (rows 4-7), the CNN and DLHD use 250-second X-ray exposures. The best performance metric for each estimated halo parameter is in bold.\label{table:DS_CNN_DLHD}}
\tablewidth{0pt}
\tablehead{
\colhead{} &
\multicolumn{3}{c}{\textbf{DS}} &
\multicolumn{3}{c}{\textbf{CNN}} &
\multicolumn{3}{c}{\textbf{DLHD}} \\
\cline{2-3}\cline{4-5}\cline{6-10}
\colhead{\textbf{Property}} &
\colhead{RMSE} & \colhead{$R^2$} & \colhead{$\chi^2$} &
\colhead{RMSE} & \colhead{$R^2$} & \colhead{$\chi^2$} &
\colhead{RMSE} & \colhead{$R^2$} & \colhead{$\chi^2$}
}
\startdata
$M_{500c}$ (250s)  & 0.104 & 0.906 & 1.209 & 0.072 & 0.955 & 0.686 & \textbf{0.056} & \textbf{0.972} & \textbf{0.911} \\
$M_{500c}$ (1ks)   & 0.104 & 0.906 & 1.209 & 0.068 & 0.960 & 0.752 & \textbf{0.060} & \textbf{0.969} & \textbf{0.864} \\
$M_{500c}$ (10ks)  & 0.104 & 0.906 & 1.209 & \textbf{0.060} & \textbf{0.971} & 0.671 & 
\textbf{0.060} & 0.970 & \textbf{1.034} \\
\hline
$f_{\rm gas}(0.5R_{200c})$ & 0.0124 & 0.791 & \textbf{0.890} & 0.0078 & 0.918 & 0.700 & \textbf{0.0071} & \textbf{0.933} & 1.413 \\
$f_{\rm gas}(0.7R_{200c})$ & 0.0155 & 0.719 & 1.104 & 0.0081 & \textbf{0.931} & 0.584 & \textbf{0.0076} & 0.928 & \textbf{1.036} \\
$f_{\rm gas}(R_{200c})$    & 0.0149 & 0.754 & \textbf{1.010} & 0.0083 & 0.913 & 0.746 & \textbf{0.0074} & \textbf{0.933} & 0.871 \\
$f_{\rm gas}(2R_{200c})$   & 0.0113 & 0.657 & \textbf{0.960} & 0.0087 & 0.796 & 0.760 & \textbf{0.0079} & \textbf{0.838} & 0.909 \\
\enddata
\end{deluxetable*}

\subsection{Inferring Gas Fractions} \label{sec:fgas}

\begin{figure*}

\includegraphics[width=\textwidth]{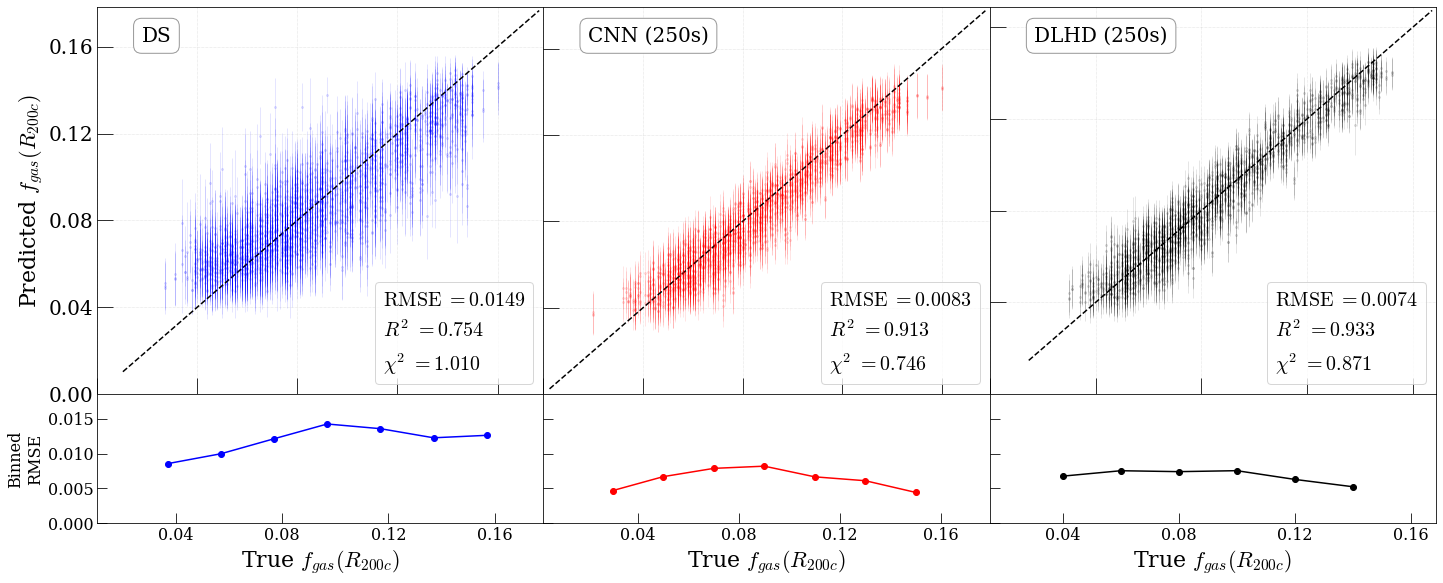}
\caption{Truth-prediction plots (top panels) and binned RMSE (bottom panels) of Deep Sets, the CNN, and the DLHD when predicting $f_{\rm gas}(R_{200c})$ on 250-second exposure X-ray mocks.}
\label{fig:comp_fgas}
\end{figure*}

We also use the DLHD to infer gas fractions ($f_{\rm gas}$), compiled by \citet{Ayromlou2023}, to determine if there is an optimal radius at which they can be measured via X-ray emission mapping. We investigate the fraction of gas contained within four different radii: $0.5 R_{200c}$, $0.7 R_{200c}$ (corresponding to $\sim R_{500c}$), $R_{200c}$, and $2 R_{200c}$. Figure~\ref{fig:comp_fgas} shows the truth-prediction plots of Deep Sets, the CNN, and the DLHD for $f_{\rm gas}(R_{200c})$, and the corresponding statistics are compiled in the bottom half of Table~\ref{table:DS_CNN_DLHD}. 

We find that the most accurately predicted gas fraction in terms of RMSE is at $0.5R_{200c}$, while accuracy generally decreases at larger radii. Further, the accuracy of DLHD outperforms its component networks across all radii, producing RMSE values $\sim 1.8\times$ smaller than those from Deep Sets and $\sim 1.1\times$ smaller than those from the CNN. We note that the Deep Sets network appears to be the best calibrated, as it produced $\chi^2$ values closer to one compared to the CNN and DLHD across most radii. However, the Deep Sets model is consistently less accurate in its $f_{gas}$ predictions, rendering its $\chi^2$ values largely uninformative. Further, the DLHD still produces favorable $\chi^2$ values, e.g. $\chi^2 = 1.036$ for $f_{gas}(0.7R_{200c})$, indicating that the DLHD is also well calibrated. 



\section{Discussion}\label{sec:discussion}

\subsection{Comparison to Previous Results}\label{sec:prevresultcomp}

Our DLHD accuracy exceeds the results of previous ML tools used to infer halo mass.  Neural networks (NNs) applied to X-ray imaging have often outperformed NNs used on galaxy catalogues, which reflects our results of the CNN outperforming the DS network.  

Regarding galaxy data, ML has been shown to outperform standard scaling metrics. Following \citet{ho_robust_2019}, a CNN-based KDE performs $2\times$ better than the $M-\sigma$ relation, reaching 0.13~dex accuracy for clusters. Further, \citet{Ma2025} developed an Artificial NN applied to galaxy catalgoues that reached 0.2~dex accuracy for groups, although they achieved better performance for TNG300. \citet{Calderon2019} demonstrated that ML algorithms can overcome bias in halo estimates using dynamical mass estimates and halo abundance matching. For comparison, our Deep Sets algorithm achieves 0.104~dex for $M_{500c}$, which is notably more accurate than the previous methods listed.  

X-ray imaging often returns better mass estimates, as shown by \citet{Ntampaka2019}, who achieved 0.05~dex accuracy with {\it Chandra} 100~ksec mocks of TNG300 clusters and massive groups.  \citet{ho_2023} reached 0.071~dex accuracy with multi-band {\it eROSITA} mocks of 2000 seconds using Magneticum mocks.  These are comparable to our CNN results of 0.060--0.072~dex across different integration lengths.  We note that \citet{Iqbal2025} achieved a precision of 0.023~dex using graph neural networks, although they used physical property profiles (density, temperature, pressure), which generally produce a higher precision than mock observations. However, in practice, these properties must be inferred and are generally not practical inputs for models, unlike the observables used in our study.  

Using multimodal networks, \citet{ho_2023} achieved only slightly better accuracy (0.070~dex) with their dual CNN architecture for halo mass inference. However, a direct comparison is complicated by differences in mock realism, in part because  \citet{ho_2023} used eROSITA-specific mocks, while the DLHD is trained on more generalized mocks detailed in \S \ref{sec:mockdatasets}. Nonetheless, the Deep Sets algorithm's ability to flexibly process unstructured datasets shows that the DLHD can significantly improve inference of halo properties in a multimodal network.  While \citet{ho_2023} stated that the inclusion of the dynamical information in a 3-D phase space cube offered little improvement, we suggest that the ability to include additional galaxy-specific properties (e.g. $M_\star$ \& SFR) leads to superior performance.  We further note that our binned RMSE at $M_{500c}>10^{13.5}\;\msolar$ is even lower than that reported for our entire sample (cf. $\sim 0.04$ vs. 0.056; lower right panel of Fig. \ref{fig:comp_m500}).  We mention this lower RMSE for the higher mass threshold since \citet{ho_2023} used a similar mass cut in their cluster sample; however, we defer a further controlled like-for-like comparison to future work. 


\subsection{DLHD Improvement upon Component Networks}\label{sec:DLHDperformance}

The DLHD achieves a combined mass scatter of 0.056~dex, outperforming the individual Deep Sets and CNN component networks. This improvement suggests that the individual networks' residual errors are partially correlated with specific physical states, and can therefore be corrected through joint analysis. If independent predictions were uncorrelated, their combination would yield accuracy comparable only to the single best-performing network. Physically, these correlated residuals represent uncertainty related to underlying astrophysical systematics. For instance, recent group mergers can cause departures from hydrostatic equilibrium, which systematically skew X-ray mass estimates \citep{sanderson_locuss_2009}. However, the dynamical information of mergers is often retained within the spatial and velocity distributions of the galaxy catalogues. By evaluating both modalities simultaneously, the DLHD utilizes learned, non-linear combinations of the data to recognize these observable signatures and correct the corresponding X-ray bias, effectively breaking observational degeneracies.

\subsection{Future Directions}
Future DLHD development will focus on scaling the architecture and transitioning to observational applications. The immediate next step is to train the network on larger catalog datasets to improve the robustness of the combined parameter inferences. We should also extend the halo mass range below the current TNG300 limit using higher-resolution simulations to better cover the low-mass group population, where AGN feedback effects on the gas fraction are strongest. Before deploying the DLHD on real datasets including DESI or {\it eROSITA} catalogs, we must systematically augment the realism of our training mocks, including DESI-specific spatial completeness and fiber-collision artifacts, as well as {\it eROSITA}'s specific signal-to-noise limitations and point-spread function, whose combined impact on the 0.056~dex $M_{500c}$ baseline established here is currently unknown. Finally, characterizing the correlated residual structure between the Deep Sets and CNN branches as a function of halo merger state and departure from hydrostatic equilibrium will clarify which physical configurations drive the multimodal performance gain and how robustly these gains will transfer from simulation to observation. Marginalizing over these survey-specific realities will enable the DLHD to provide reliable, combined stellar and gas mass predictions for ongoing and upcoming galaxy surveys.
\section{Conclusions}\label{sec:conclusion}

We develop a novel, multi-modal neural network called the Deep Learning Halo Definer (DLHD) that inputs both galaxy catalogue and image-based datasets to more accurately return halo masses and halo gas fractions. We used the IllustrisTNG300 simulation to create sets of realistic groups and low-mass clusters.  For the catalogue datasets, we use the Deep Sets (DS) neural network algorithm \citep{zaheer_deep_2017} and test it on different data quality cuts corresponding to potential observational uncertainties in real data. We also used a CNN on mock X-ray images of varying exposure lengths. First, exploring the individual DS and CNN networks on their own, we find the following:

\begin{itemize}

    \item We suggest that the DS algorithm's ability to flexibly process unstructured datasets that include galaxy-specific properties, such as stellar mass, achieves improved metrics over previous multimodal network results. In particular, the DS network is able to infer halo masses with an RMSE of $0.104$~dex when including idealized galaxy data down to $10^9\;\msolar$.

    \item Setting a higher stellar mass threshold cut does not significantly degrade the DS inference, as the central galaxy is still identified and centered. However, uncertainty in identifying the center of the group, as well as the addition of velocity uncertainties $\ga 100\;\kms$, significantly degrades the DS inference.  

    \item The CNN consistently outperforms the DS network, reaching a halo mass inference RMSE of 0.072~dex for 250-second exposures, similar to the {\it eROSITA} eRASS:1 survey depth, but not significantly improving for longer X-ray exposures.  
    
\end{itemize}

The DLHD network combines the latent vector outputs of the DS and CNN networks and outperforms both networks individually. We find:
\begin{itemize}
    \item The DLHD inference of halo mass on 250~sec exposures achieves an RMSE of 0.056~dex, representing a factor of $1.8\times$ improvement over the DS network and $1.3\times$ over the CNN. 
    \item The DLHD also achieves superior inference for halo gas fractions across different radii. Using DLHD, the precision of $f_{\rm gas}$ within $R_{200c}$ has an RMSE of 0.0071, representing a factor of $2.0 \times$ improvement over the DS network and $1.1 \times$ over the CNN.  

\end{itemize}

\section*{Acknowledgements}
We acknowledge and thank Michelle Ntampaka and Helen Shao for their advice and discussions throughout this project. Mohammadreza Ayromlou is supported at the Argelander Institut für Astronomie through the Argelander Fellowship. This work utilized the Alpine high-performance cluster at the University of Colorado Boulder and the Mortimer Research Cluster at the University of Wisconsin Milwaukee. This work is supported by National Science Foundation grants AST-2206055, AST-2511137 and PHY-2513124.

\section*{Data Availability}
\label{sec:data avail}
The IllustrisTNG300 simulation data underlying this work are publicly available at www.tng-project.org. The DLHD code, Optuna databases, Pytorch configurations, and trial results are available on GitHub at \href{https://github.com/CalebBOgle/DLHD}{https://github.com/CalebBOgle/DLHD}.

\vspace{5mm}

\bibliography{main}{}
\bibliographystyle{aasjournal}

\end{document}